# Can an Agency Role-Reversal Lead to an Organizational Collapse?; A Study Proposal

Yossi Haimberg                                                                                                      May 10, 2021

Portland State University

## Abstract

The Principal-Agent Theory model is widely used to explain governance role where there is a separation of ownership and control, as it defines clear boundaries between governance and executives. However, examination of recent corporate failure reveals the concerning contribution of the Board of Directors to such failures, and calls into question governance effectiveness in the presence of a powerful and charismatic CEO. This study proposes a framework for analyzing the relationship between the Board of Directors and the CEO, and how certain relationships affects the power structure and behavior of the Board, that leads to a role reversal in the Principal-Agent Theory, as the Board assumes the role of the CEO's agent. This study's results may help create "red-flag" for a board and leader's behavior that may result in governance failure.

**A.    Background**

Seismic board failures emphasize the need for a nuanced understanding of a BOD's (Board of Directors) social and psychological dynamics to evaluate, and ultimately change, the conditions that may lead to such failures. Under agency theory, the comprehensive theory of the firm assumes separation of ownership (principal) and control (agents) in a hierarchical structure that is both monitored and incentivized to induce managers to act in the best interest of the stakeholder (Jensen and Meckling, 1976). Organizations such as private, public, and non-profit corporations have multi-level principal-agent arrangements in which governance is assigned to the board, which problematically acts as both agent of the stakeholders and principal over the executive management. Wiseman et al. (2012) and Van Ees and Huse (2009) suggest that the BOD controls an organization in name only as leaders[1] control and influence board behavior through agenda-setting, the flow of information to the board, and hiring/firing powers, giving the leader complete control over the organization. Yet power and influence are key aspects of behavior in all organizations (Ott et al. 2008). When a BOD relinquishes its powers, voluntarily or under threats, to the leader's authority, its behavior tends to deteriorate towards non-fiduciary and complacency to avoid cognitive conflict within the BOD or getting fired by the leader.

   I posit that the consequences of power consolidation by a leader over the BOD triggers a principal-agent role reversal (dependent variable), commonly understood as a structural breakdown in the traditional, hierarchical principal-agent configuration, adversely affecting organizational governance. Board members who are now motivated by coercion, fear and greed[2], betray their legal and fiduciary responsibilities and allow the leader to act as she pleases, sometimes with catastrophic results. The questions this research will address are: Why does the Board of Directors (BOD) and executive power dynamic flip?, How does executive capture of BOD influence BOD decisions?, and haw does agency role reversal affect the organization?. Appendix "A" depicts the proposed research outline, the independent variables that cause role reversal, and the proposed methodology.

### B.   Literature review and Theoretical Background

The principal-agent theory provides a theoretical framework for examining the multitude of relationships in an organization (Jensen and Meckling, 1976, Fama and Jensen 1983). The BOD is a fiduciary agent of the stakeholders with an end goal of overseeing the successful, profitable, and sustainable operations of an organization (Rosenblum et al. 2014). The BOD is entrusted with significant governance powers to act vis-à-vis management, through incentives and controls, to motivate a leader to act in the best interests of the stakeholders. Notwithstanding this simple hierarchical check-and-balance structure motivated by mostly rational and utility-maximizing behavior through economic incentives, scholars suggest the need for behavioral processes and dynamics in and around the BOD for a better understanding of the conditions that affect governance (Forbes and Milliken, 1999, McNulty and Pettigrew, 1999, Westphal, 1999). Factors such as judgement, decision making and behavior are all affected by heuristics and individual cognitive biases (Tversky and Kahneman, 1974). Pressure towards conformity and cohesion within groups (Asch, 1956, Janis, 1982), and exogenous factors such as consumers, media, auditors, laws and regulations, and management's influence over the BOD (through leader's board directorship and Chairmanship) can all lead to conflict and political maneuvering around social power and influence in an organization (Ott et al., 2008, p. 338). French and Bertram (1959, p. 339) provide five bases of social power and influence: reward power, the perception of coercive power, organizational authority, referent power, and expert power. These bases provide the framework for assessing the question of who controls the BOD.

---

1. I use the term "leader" as a general term to denote titles that vary between organization: CEO, President, Executive Director etc.
2. Average director's 2020 compensation in large organizations was $290,000 (FW Cock, 2020)

**Reward (punishment) power,** one of the most important tasks of a BOD, is the selection, recruitment and monitoring of an organizational leader in which the BOD is expected to act as a principal. However, once a leader is installed, a "capture" of the BOD starts (Main et al., 1995) and the leader consolidates power and influence through her authority to choose and recommend members for the BOD approval (Lorsch, 1989).

**The perception of coercive power** demonstrates that, in addition to appointment or removal powers of independent directors, the leader also nominates and appoints members of her management team to the BOD, over which she has even more power since their disloyalty to her can cost them their jobs. The perceived threat of a director's removal allows leaders to promote corporate culture and a board's "cohesiveness," consolidating authority, and potentially turning the BOD into a leader's rubber stamp.

**Organizational Authority**: It is important to note that power consolidation is possible only because institutional investors who control 70% of corporate stocks are inclined to shy away from BOD participation because of regulatory and fiscal rules, such as diversification, and interpretations of the insider trading regulations (Tirole, 2001). Once a leader's duality is in place (leader and Chairman), leaders complete their power consolidation and achieve de-facto legitimacy of authority. Alexander and Weiner (1998) note a similar situation exists in nonprofit corporations.

**Referent power** indicates that leaders may possess skills, experience and reputations that make them particularly valuable as outside directors of other firm's BOD (Adams and Ferreira, 2007). Hermalin and Weisbach (1998) note that the more successful and powerful the CEO, the more likely she will be reappointed to other boards. In addition to reputation building, such appointments allow information gathering, expanded social and professional personal ties, and added prestige (French and Raven, 1959).

**Expert powers** indicate that both the leader and management have superior knowledge and complete operational and financial information. They have control over the flow, interpretation and timing of the organizational, operational, and financial information given to the BOD. Information asymmetry, at the heart of the principal-agent theory, demonstrates how power and influence affect an organization.

The framework noted above can lead to tyranny in organizations (Bies and Tripp, 1998), and to shocking and outrageous abuses of power, epitomized by Tyco partially funding a $2.1 million birthday party in 2002 for the wife of Chief Executive Officer Dennis Kozlowski, or United Way funding the lavish lifestyle and extramarital affair of its president, William Aramony. Central to these stories is the assumption that somehow governance is to blame—that is, the system of checks and balances meant to prevent abuse by leaders (Larker and Tayan, 2011). Hagberg (2003) and Kanter (1979) suggest that the underrepresentation of women in board's and other organizational echelons harms performance and behavior. Women can act as counterbalance to a board's aggressive, competitive social dynamics, which is "gendered" behavior that may contribute to organizational failure.

C.    **Model, and Statement of the model hypothesis**

The board plays a central role in the agency model of an organization, acting in a dual capacity as both principal and agent. Yet the intended hierarchical structure of stakeholders-corporate governance by the BOD-leader is affected by the leader's behavior towards power consolidation, resulting in a "captive" BOD, controlled by the leader, who now controls the organization. My research question explores this problematic dynamic. What are the factors that can explain this power transition, and the resultant social influence on directors' behavior that lead to agency role reversal? Because of confidentiality restrictions on the inner workings of a board's behavior and relations, this study is limited to worse-scenario, failed organizations where data is more readily available. I propose a collective case study that involves simultaneously analyzing multiple cases of US companies and nonprofit organizations such an Enron, Theranos, AIG, and a non-profit, such as the defunct UnitedWay. The case study is based upon the following hypotheses:

1. Charisma, reputation and tenure play a significant role in the selection and appointment of leaders in failed organizations;
2. Leaders' capture of BOD happens in accordance with French and Bertram's (1959) framework;
3. Captured boards tend to be less sensitive or responsive to red flags;
4. Captured boards are more cohesive, compliant, and problematically, less diversified;
5. Organizational complexity and bounded rationality reduce board members' ability to control the leader;
6. Failed organizations have a lower ratio of female to male directors on a BOD.

D.   **An outline of possible methodology and methods for the research design**

I propose the following methodology for understanding the power capture by a leader in social organizations.

1. Examine any investigative materials from the Securities and Exchange Commission (SEC) and other investigations (law enforcement, Senate).
2. Collect data on the leader (tenure, reputation, experience), board structure (diversity, committees, size, ratio of outside to inside directors), board cohesiveness measures (composition: gender, race, age, occupation, status, stability), distinction (how many other boards), uniformity (habitual routines), norms (director's participation, full-board discussion limits, structure, formalities, conformity, courage to challenge the leader), and voting records.
3. Search for commonalities between the firms in the study that indicate power accumulation and damaging role reversal.,
4. Make observations based on the theoretical contexts presented above and suggest recommendations based on the gathered evidence.

E.  **Contribution to Literature**: This research will contribute to organizational behavior within boards in two ways. First, by using organizational behavioral theory, I suggest "red-flags" for a board and leader's behavior that may result in governance failure. Second, I purport that organizational behavior theory can affect changes within the principal-agent model that result in agency role reversal.

The novel concept of agency reversal is a clear breakaway from the hierarchical agency model, and its contribution to agency studies can be significant, particularly in the field of public administration and policy studies where the agency model is widely used to explain multi-level, hierarchical organizational structures and dynamics, but not without considerable limitations. Moe (1984) looks at how bureaucratic superiors can control their subordinates and notes that a bureaucratic motivational structure that focuses on security, career opportunities, policy, budget and slack, and a complex, inflexible distributed bureaucratic hire and fire authority undermines agency theory's organizational control. Within the agency context, bureaucratic superiors have limited control over subordinates, and limited capacity for incentives. He also notes a politician's control over the bureaucrat in a multi-level agency structure where multiple actors serve as both principal and agent simultaneously, and where seemingly unbridgeable interests and monitoring mechanisms are present, results in the absence of these rigid control, monitoring and incentive systems. The agency model in the public sector falls short in explaining these relationships, or even in suggesting how to align a bureaucracies' interests with the politicians.'

Yet, agency role reversal can explain the power struggle between politically appointed, short-term (election cycle or before) leaders of government agencies who are guided by high-level political aspirations, and the well-entrenched bureaucratic organization that existed before the new appointee took charge and will continue existing after his replacement. The information asymmetries could not be more distinct, and without effective monitoring, the bureaucrat is the tail that wags the dog.

With these theoretical and relational contexts in mind, agency role reversal is an exciting, novel addition to the agency theory that is not present in the current agency theory literature. Identifying agency reversal situations is important; it indicates a breakdown in the intended organizational structure and purpose, often with governance inefficiencies that result in severe and costly consequences.

# Appendix "A"

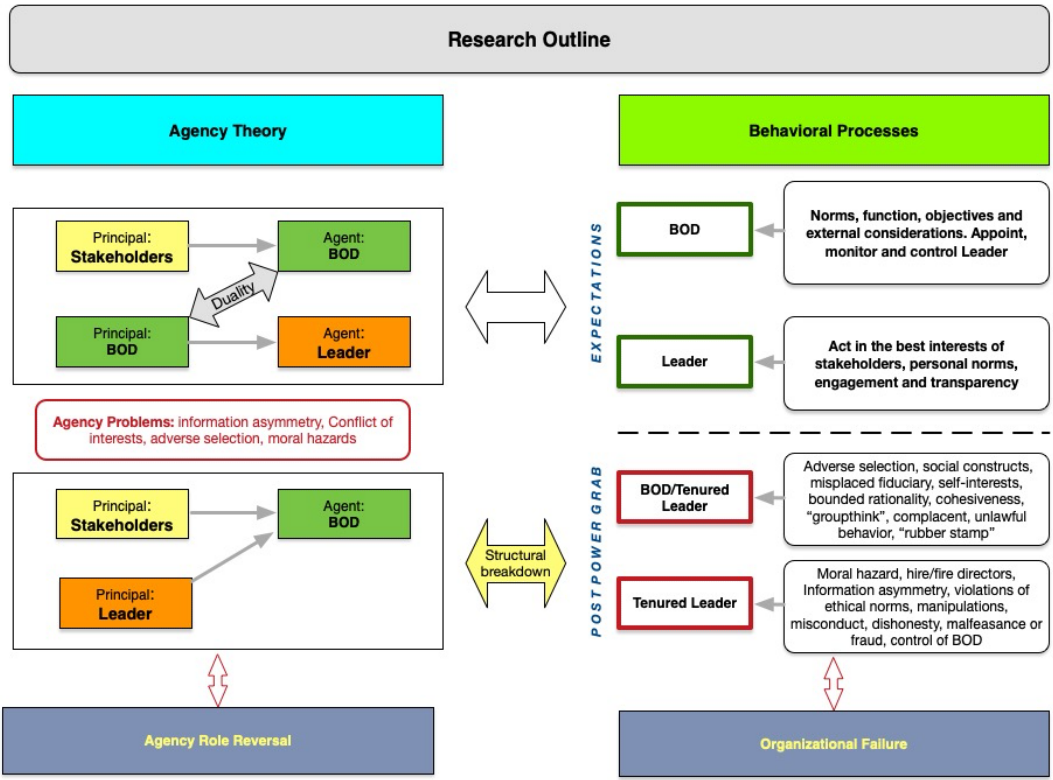

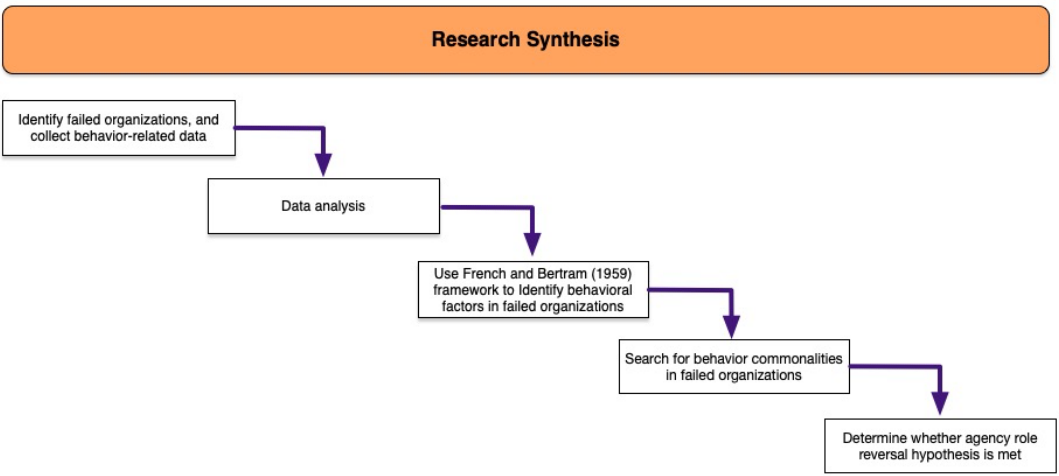

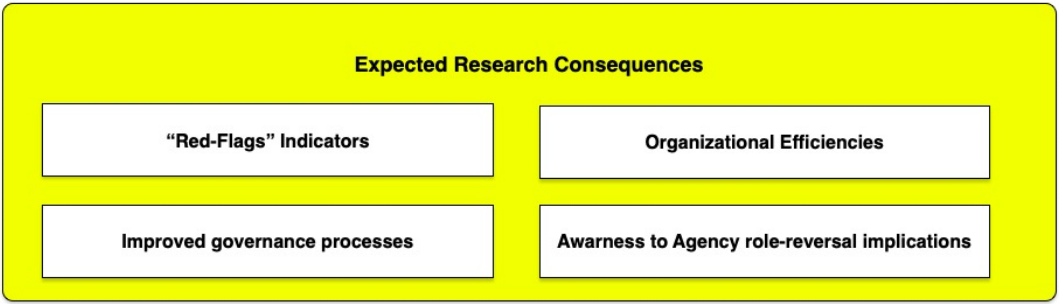

# References

Adams, R. B., and Ferreira, D. (2008). Do directors perform for pay? Journal of Accounting and Economics, 46(1), 154–171. https://doi.org/10.1016/ j.jacceco.2008.06.002. http://www.sciencedirect.com/science/article/pii/ S0165410108000347

Alexander, J. A., & Weiner, B. J. (1998). The adoption of the corporate governance model by non- profit organizations. *Nonprofit Management & Leadership*, 8, 223-242.

Asch, S.E. (1956). Studies of Independence and Conformity: A Minority of One Against a Unanimous Majority. Psychological Monographs 70(9).

Bies R.J. and Tripp, T.M. (1998), Two Faces of the Powerless: Coping with Tyranny in Organizations, in Ott, J.S., Parkes, S.J. and Simpson, R.B., (2008), Classic Readings in Organizational Behavior, 4th Edition, Wadsworth Cengage Learning, 2008.

Fama, E. F., & Jensen, M. C. (1983). Separation of ownership and control. Journal of Law and Economics, 26, 301–325.

Forbes, D.P. and F.J. Milliken: 1999, "Cognition and corporate governance: Understanding boards of directors as strategic decision making groups", Academy of Management Review 24: 489-505.

French, J.R.P. and Bertram R. (1959), The Base of Social Power, Institute of Social Research, in Ott, J.S., Parkes, S.J. and Simpson, R.B., (2008), Classic Readings in Organizational Behavior, 4th Edition, Wadsworth Cengage Learning, 2008.

FW Cook (2020), 2020 Director Compensation Report, FW Cook.

Hagberg, J. O. (2003), Women in Power, in Ott, J.S., Parkes, S.J. and Simpson, R.B., (2008), Classic Readings in Organizational Behavior, 4th Edition, Wadsworth Cengage Learning, 2008.

Hermalin, B. E., and Weisbach, M. S. (1998). Endogenously chosen boards of directors and their monitoring of the CEO. The American Economic Review, 88(1), 96–118; http://www.jstor.org/stable/116820

Janis, Irving (1982). Groupthink. Second Edition. Boston: Houghton Mifflin.

Jensen, M.C. and Meckling, W.H. (1976), Theory of The Firm Managerial Behavior, Agency Cost and Ownership Structure, Journal of Financial Economics 3 (1976) 305-360. Q North-Holland Publishing Company.

Kanter, R.M. (1979), Power Failure in Management Circuits Harvard Business School Publishing, in Ott, J.S., Parkes, S.J. and Simpson, R.B., (2008), Classic Readings in Organizational Behavior, 4th Edition, Wadsworth Cengage Learning, 2008.

Larkin, D.L. and Tayan, B. (2011), Corporate Governance Matters: A Closer Look at Organizational Choices and Their Consequences, Pearson Education, Inc. 2011.

Lorsch, J. (1989), Pawns or Potentates: The Reality of America's Boards. Harvard Business School Press: Boston, MA.

McNulty, T. and A. Pettigrew, (1999), Strategists on the board, Organization Studies 20: 40-74.

Moe, T.M. (1984), The New economics of Organization, American Journal of Political Science, Vol. 28, No. 4 (Nov., 1984), pp. 739-777.

Ott, J.S., Parkes, S.J. and Simpson, R.B., (2008), Classic Readings in Organizational Behavior, 4th Edition, Wadsworth Cengage Learning, 2008.

Rosenblum, S.A., Cain, K.A., and Niles, S.V., (2014), NYSE: Corporate Governance Guide, White Page Ltd.

Tirole, J. (2001), 'Corporate Governance Econometrics', Econometrica, Vol. 69, No. 1., Jan., 2001, pp. 1-35.

Tversky, Amos; Kahneman, Daniel (1974). " Judgement under uncertainty: heuristics and biases." Science 184, 1124-1131.

Van Ees, H. and Huse, M. (2009), Toward a behavioral theory of boards and corporate governance, Corporate Governance: An International Review (2009, 17 (3): 307-319.

Westphal, J.D.: 1999, "Collaboration in the boardroom: behavioural and performance consequences of CEO-board social ties", Academy of Management Journal 42: 7-24.

Wiseman, R.M., Cuevas-Rodriguez, G. and Gomez-Mejia, L.R., (2012), Towards a Social Theory of Agency, Journal of Management Studies, 49:1 January 2012, doi: 10.1111/j.1467-6486.2011.01016.x